\documentclass[twocolumn,aps,prl,epsf,superscriptaddress,showpacs]{revtex4}
\usepackage{amsmath}
\usepackage{epsfig}
\usepackage{epsf}
\usepackage{array}

\begin{document}

\title{Lattice relaxation, electronic screening, and spin and orbital phase diagram of 
LaTiO$_3$/SrTiO$_3$ superlattices}
\author{Satoshi Okamoto}
\email{okapon@phys.columbia.edu}
\affiliation{Department of Physics, Columbia University, 538 West 120th Street, New York,
New York 10027, USA}
\author{Andrew J. Millis}
\affiliation{Department of Physics, Columbia University, 538 West 120th Street, New York,
New York 10027, USA}
\author{Nicola A. Spaldin}
\affiliation{Materials Research Laboratory and Materials Department, 
University of California, Santa Barbara, California, 93106, USA}
\date{\today }

\begin{abstract}
The effects of lattice relaxation in LaTiO$_3$/SrTiO$_3$ superlattices are investigated 
using a combination of LDA$+U$ density functional theory, and Hartree-Fock 
effective Hamiltonian calculations. 
We find noticeable ($\sim 0.1$--0.2~\AA) distortions of the TiO$_6$ octahedra in the near-La region. 
The resulting screening changes the Ti $d$-electron density substantially. 
Tight-binding fits to the relaxed-lattice band structure, combined with Hartree-Fock 
calculations of the resulting model, reveal a novel phase with $xy$ orbital order, 
which does not occur in bulk LaTiO$_3$, or in the hypothetical unrelaxed structure. 
\end{abstract}


\maketitle


Recent advances in the techniques of pulsed laser deposition and molecular beam epitaxy have allowed 
the creation of ``oxide heterostructures'' consisting of alternating layers (of arbitrary number of unit cells) 
of different transition metal oxide compounds. 
The physics becomes particularly interesting when one or more of the constituent layers is a compound with 
correlated electron properties \cite{Imada98,Tokura00} such as Mott insulating behavior, 
high temperature superconductivity or colossal magnetoresistance. 
In addition to advances in fabrication,
recent years have seen tremendous progress in the characterization 
\cite{Izumi01,Ohtomo02,Gariglio02,Bozovic03,Yamada04,Varela04,Schneider04,Pena05,Stahn05,Logvenov05,Kavach05}
and theoretical analysis
\cite{Okamoto04a,Okamoto04b,Okamoto05,Popovic05,Freericks04,Pavlenko05} of such structures. 
However, before this work, the crucial issue of lattice relaxation has not been systematically addressed. 

We expect the lattice relaxation issue to be important because a key feature of heterostructures is 
``charge redistribution'' \cite{Ohtomo02}; that is the redistribution of electrons from one layer to 
another, and consequent changes from the bulk valence values, 
driven by the difference in electrochemical potentials between the component materials. 
Simple estimates and tight-binding model studies suggest that the resulting dipole layer involves 
a charge density of about $\frac{1}{2}e/\rm unit \, cell$, displaced by one to two lattice constants 
\cite{Okamoto04a}. 
The electric field associated with such a dipole layer is not small, $\sim 0.1$~eV/\AA, 
and may be expected to drive significant changes in atomic positions compared with the
bulk materials. 
Further, many experimentally relevant heterostructures involve ferroelectric or nearly ferroelectric materials, 
for example SrTiO$_3$, for which enhanced dielectric response is an issue. 

The purpose of this paper is to calculate the magnitude and nature of the lattice 
relaxation in LaTiO$_3$/SrTiO$_3$ heterostructures, and to determine its effect on 
the electronic properties. First, we calculate the relaxed structures of 
[LaTiO$_3$]$_n$[SrTiO$_3$]$_m$ superlattices using the LDA$+U$ method of density functional 
theory within the projector augmented wave (PAW) approach \cite{Bolechl94} as implemented in 
the {\it Vienna Ab initio Simulation Package} (VASP) \cite{Kresse96,Kresse99}. 
We consider [001] ($n$-$m$) superlattices in which a unit consisting of $m$ planes 
of LaTiO$_3$ followed by $n$ planes of SrTiO$_3$ layers is repeated in the [001] ($z$)
direction, similar to those studied experimentally by Ohtomo {\it et al}. \cite{Ohtomo02}.
Technical details include the rotationally invariant LDA$+U$ method of Liechtenstein {\it et al}. 
\cite{Liechtenstein95} with $U=5$ and $J=0.64$~eV for the Ti $d$ states \cite{Mizokawa95}.  
In addition, 
if we treat the La $f$ states within the LDA, we find that the empty La $f$ bands lie only $\sim 2$~eV 
above the Fermi level, leading to a spurious mixing and level repulsion with the Ti-derived $d$ bands. 
Since in practice, the La $f$ bands should lie much higher in energy \cite{Czyzyk94}, we impose a large 
$U$ of 11 eV, and $J=0.68$~eV on the La $f$ states. Omitting the La $f$ $U$ changes the results, 
for example reducing the lattice distortions by $\approx 50$~\%. 
For Sr and Ti, we use PAW potentials in which semi-core $s$ states are treated as valence states, 
(Sr$_{sv}$ and Ti$_{sv}$ in the VASP distribution) 
while for La and O, we use standard potentials (La and O in the VASP distribution), and 
we use a $4 \times 4 \times 1$ $k$-point grid and an  energy cutoff 500~eV. 
The lattice constants $a$ and $b$ are fixed to the experimental value for
cubic SrTiO$_3$ (3.91~\AA) which is the substrate used in the experiments \cite{Ohtomo02}. 
The $c$ axis lattice constant and atomic $z$ coordinates are adjusted, while retaining the 
tetragonal symmetry of the crystal, until the forces on the ions are less than 0.01 eV/\AA.

\begin{figure}[tbp]
\includegraphics[width=1\columnwidth,clip]{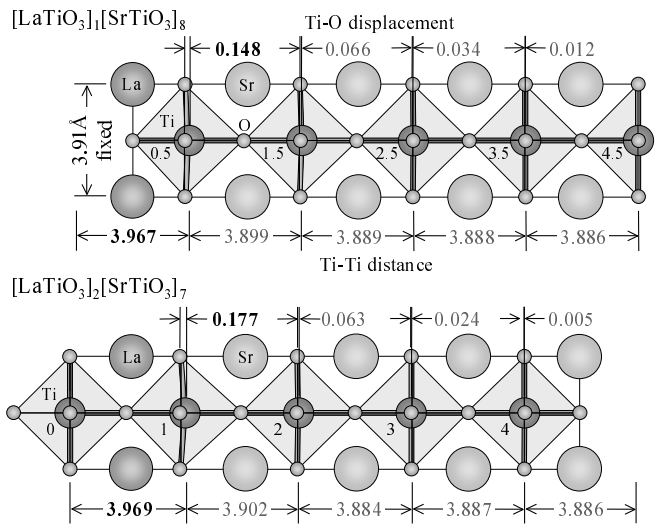}
\caption{Calculated optimized lattice structures of superlattices 
[LaTiO$_3$]$_1$[SrTiO$_3$]$_8$ (upper figure) and [LaTiO$_3$]$_2$[SrTiO$_3$]$_7$ 
(lower figure); half of the unit cell is shown in each case.
The optimized $z$ axis lattice constants are $c=35.09$~\AA\, and 35.17~\AA \, for 
[LaTiO$_3$]$_1$[SrTiO$_3$]$_8$ and [LaTiO$_3$]$_2$[SrTiO$_3$]$_7$, respectively. 
The intertitanium distances (lower lines) and displacements of the Ti ions relative 
to the O$_2$ planes (upper lines) are also indicated. 
The center of the LaTiO$_3$ region is taken as the zero of the $z$ coordinate
in each case and the Ti ions are labeled by their relative $z$ positions. 
}
\label{fig:structure}
\end{figure}

Figure~\ref{fig:structure} shows our calculated relaxed lattice structures for two representative 
cases: (1-8) and (2-7) heterostructures. The largest structural relaxations occur in 
the TiO$_2$ layer at the LaTiO$_3$-SrTiO$_3$ interface (Ti$_{0.5}$ in the upper panel and Ti$_1$ 
in the lower panel) with the Ti displaced from its ideal position by 0.15~\AA \, in the (1-8) case 
and 0.18~\AA \, in the (2-7) structure. As a consequence, the lengths of the Ti-Ti distances across 
the LaO planes (Ti$_{-0.5}$-Ti$_{0.5}$ and Ti$_0$-Ti$_{\pm 1}$) are approximately 2~\% larger than
those across SrO planes. This ``ferroelectric-like'' distortion produces a local ionic 
dipole moment which screens the Coulomb field created by the substitution of
Sr$^{2+}$ by La$^{3+}$ ions. Moving further away from the interface, the magnitude of the 
ferroelectric-like distortion decays rapidly, while the Ti-Ti distance reverts to a constant 
value very close to that in bulk SrTiO$_3$. 

\begin{figure}[tbp]
\includegraphics[width=0.8\columnwidth,clip]{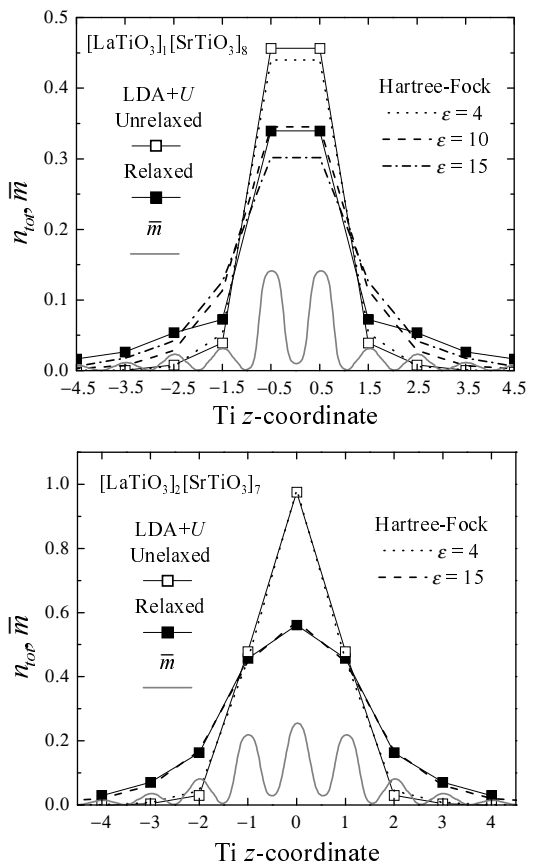}
\caption{Charge and magnetization densities for (1-8) and (2-7) heterostructures. 
Filled and open squares: conduction band charge densities per unit cell, 
relaxed and unrelaxed superlattices, respectively, obtained as described in test. 
Light lines: smoothed magnetization densities of relaxed heterostructures. 
Dotted, broken, and dash-dot lines: results of model Hartree-Fock calculation with realistic band parameters 
for unrelaxed and relaxed structures. 
$U=5$ and $J=0.64$~eV, and dielectric constant $\varepsilon$ indicated. 
}
\label{fig:MandN}
\end{figure}

An important quantity for physical insight and more detailed theoretical analysis is the spatially 
resolved conduction band charge density: loosely speaking, the Ti $d$ occupancy. To obtain this, we 
make use of the fact that the ground state within LDA+$U$ is a highly polarized ferromagnetic state  
in which the magnetization density can be ascribed to the conduction bands. Following Ref.~\cite{Sai05}, 
we compute a smoothed magnetization density $\overline m(z)$, 
shown as the light gray line in Fig.~\ref{fig:MandN}, 
by planar averaging and smoothing in $z$ over a range $\pm a/2$. 
We identify the integral of $\overline m(z)$ over a unit cell 
with the conduction-band charge density in that cell. The total (summed over all cells) conduction-band 
charge obtained in this way is within $\sim 1$~\% of the expected 1 electron per La ion in the two-layer 
heterostructure, but only $\sim$ 0.85 electrons per La in the one-layer structure 
probably because the one-layer structure is not fully spin-polarized. 
Therefore in the latter case we renormalize the density appropriately. 

Fig.~\ref{fig:MandN} compares our calculated Ti $d$ charge densities for relaxed 
(black squares) and unrelaxed (white squares) lattices 
for [LaTiO$_3$]$_1$[SrTiO$_3$]$_8$ (upper panel) 
and [LaTiO$_3$]$_2$[SrTiO$_3$]$_7$ (lower panel). 
In both superlattices, 
the screening provided by lattice relaxation reduces the charge density on the central Ti layer 
and produces a long ``tail'' in the charge distribution, extending far away from the interface.
The effect is particularly large in the two-layer structure, reducing the middle-layer density by 
almost a factor of two.
We note that in each case the interface layer (Ti$_{0.5}$ or Ti$_1$) 
remains electronically well defined, with the density dropping by approximately 0.3 electrons
between the Ti at the interface, and its neighbor surrounded by two SrO layers. 
The relaxed-lattice charge densities agree within experimental uncertainties 
with the Ti$^{3+}$ values measured by Ohtomo {\it et al}.\cite{Ohtomo02},
both in terms of peak values (experiment: 0.3 in one-layer and 0.4 in two-layer) 
and of the slow decay away from the central region. 

In addition to changing the charge-density profile, as shown in Fig.~\ref{fig:MandN}, the changes in
interatomic distances associated with the lattice relaxation lead to changes in the orbital overlaps. 
To address the effect of these changes in electronic structure on the many-body physics, we use the 
LDA$+U$ results to derive a tight-binding model, which we solve in the Hartree-Fock approximation, 
computing relaxed and unrelaxed heterostructures. 
The form of the tight-binding model is discussed in Ref.~\cite{Okamoto04a}. 
It has three classes of parameters: 
the electronic structure, parameterized by level splittings and hoppings, which we obtain by fitting to 
our LDA$+U$ calculations, 
``Kanamori'' multiplet interactions, which we treat as adjustable parameters, 
and the screening of the long-ranged Coulomb interaction, which we parameterize by a dielectric constant 
$\varepsilon$ chosen to approximately reproduce the charge densities of Fig.~\ref{fig:MandN}. 

Following previous work \cite{Okamoto04a}, 
we represent the electronic structure by a tight-binding model involving three
orbitals labeled $\alpha=xy, xz, yz$ representing three $t_{2g}$-symmetry Ti-O antibonding bands, 
with on-site energies $\varepsilon_\alpha$ and nearest-neighbor hopping $t_\alpha^\delta$ 
along $\delta=x,y,z$ directions 
given by the usual Slater-Koster rules. (further-neighbor hoppings are factors of 5-10 smaller) 
The electronic-structure Hamiltonian is 
$H_{el}=\sum_l[H_\parallel(z_l)+H_\perp(z_l)]$ with 
\begin{eqnarray}
H_\parallel (z_l) \!\! &=& \!\! 
\sum_{\alpha, \vec k_\parallel} 
\bigl[-2 t_\alpha^x (z_l) \cos k_x - 2 t_\alpha^y (z_l) \cos k_y + \varepsilon_\alpha (z_l) \bigr] \nonumber \\
&& \times d_{\alpha \vec k_\parallel}^\dag (z_l) d_{\alpha \vec k_\parallel} (z_l),  \\
H_\perp (z_l) \!\! &=& \!\! 
\sum_{\alpha, \vec k_\parallel} \Bigl[ - t_\alpha^z (z_l) 
d_{\alpha \vec k_\parallel}^\dag (z_l) d_{\alpha \vec k_\parallel} (z_l+1) + H.c.
\Bigr]. 
\end{eqnarray}
The heterostructures shown in Fig.~\ref{fig:structure} have too many bands for a direct tight-binding analysis 
to be practical. We observe that there are three kinds of Ti site in the heterostructures 
(those surrounded by La, surrounded by Sr, or with two La and two Sr neighbors), 
and correspondingly three types of Ti-Ti bonds. 
Since the hopping parameters depend on the Ti environment, we obtain them by fitting bands  obtained 
from LDA+U calculations for simplified two-layer heterostructures with atomic positions fixed to those 
found in the full (1-8 and 2-7) structures. 
Hopping parameters between different kinds of Ti along $z$ direction in the heterostructures are estimated 
considering the fact that the hoppings are in fact ``second order'' processes via O $p$ state. 
During these additional band structure calculations, we eliminate the on-site Coulomb interaction for 
Ti $d$ states which may cause an additional level splitting among $d$ states, 
while the on-site Coulomb interaction for La $f$ states is preserved. 

\begin{table}[tbp]
\caption{Tight-binding parameters for symmetry-inequivalent hoppings in units of eV derived 
by fits to LDA+$U$ band calculations of simplified heterostructures with atomic positions 
taken from LDA+U calculations of relaxed and unrelaxed 
[LaTiO$_3$]$_1$[SrTiO$_3$]$_8$ (labeled 1) and [LaTiO$_3$]$_2$[SrTiO$_3$]$_7$ (labeled 2) heterostructures. 
Parameters of relaxed (R) and unrelaxed (U) heterostructures are shown in upper and lower lines, respectively. 
For larger $z_l$, 
$t_{xy}^{x,y}=t_{xz}^{x,z}=t_{yz}^{y,z}=0.5$~eV 
and $\varepsilon_{xy,xz,yz}=0$.} 
\label{tab:parameter}
\begin{center}
\begin{tabular}{c|cc|cc|cc|cc|cc} \hline 
      & $t_{xy}^{x,y}$ & & $t_{xz}^x$ & & $t_{xz}^z$ & & $\varepsilon_{xy}$ & & $\varepsilon_{xz}$& \\
\hline \hline
$z_l$ &  -- & 0.5 & -- & 0.5  &$-0.5$& 0.5 & -- & 0.5 & -- & 0.5 \\
1 R  &  -- & 0.53 & -- & 0.46 & 0.44 & 0.57 & -- & 0.23 & -- & 0.19 \\
1 U  &  -- & 0.55 & -- & 0.55 & 0.56 & 0.51 & -- & 0.27 & -- & 0.36 \\
$z_l$ &  0  &  1   &  0 &   1  &  0   &  1   &  0 &  1   &  0 &  1   \\ 
2 R  & 0.63& 0.53 &0.63& 0.38 & 0.47 & 0.58 &0.66& 0.06 &0.66& $-0.07$ \\
2 U  & 0.63& 0.55 &0.63& 0.55 & 0.59 & 0.51 &0.66& 0.27 &0.66& $ 0.36$ \\
\hline
\end{tabular}
\end{center}
\end{table}

Representative values for $t_\alpha^\delta$ and $\varepsilon_\alpha$ are given in table~\ref{tab:parameter}. 
We see that the largest effect of the relaxations is a decrease in the hoppings 
for $d_{xz}$ and $d_{yz}$ orbitals at the interface layers, 
and that the larger distortions in the two-layer heterostructure lead to larger effects. 
Decrease in the in-plane hoppings for $d_{xz}$ and $d_{yz}$ is 
as large as $\approx 30$~\% in the (2-7) heterostructure and that in the $z$-direction hoppings is 
$\approx 20$~\% in both the (1-8) and (2-7) heterostructures. 
Similarly, a moderate level splitting ($\sim 0.13$~eV) occurs in the transition layer. 

To extract appropriate values for the dielectric constant, $\varepsilon$, we add the long-ranged Coulomb 
terms to the tight-binding model (for details see Ref.~\cite{Okamoto04a}), and adjust $\varepsilon$ to 
fit the LDA$+U$ charge densities, as shown in Fig.~\ref{fig:MandN}. 
One sees that the results for the unrelaxed case are well
described by a $\varepsilon$ of 4, and that the effects of screening from the lattice relaxations can be 
simulated reasonably well (but not perfectly) by increasing the dielectric constant to $\approx 15$. 
The main deficiency is that the dielectric constant model overestimates the rate at which charge 
density decays far from the heterostructure, in the $n \alt 0.05$ region, for the (1-8) heterostructure 
(cf. upper panel). 
We associate this with the nonlinear screening characteristic of nearly ferroelectric materials. 

\begin{figure}[tbp]
\includegraphics[width=0.8\columnwidth,clip]{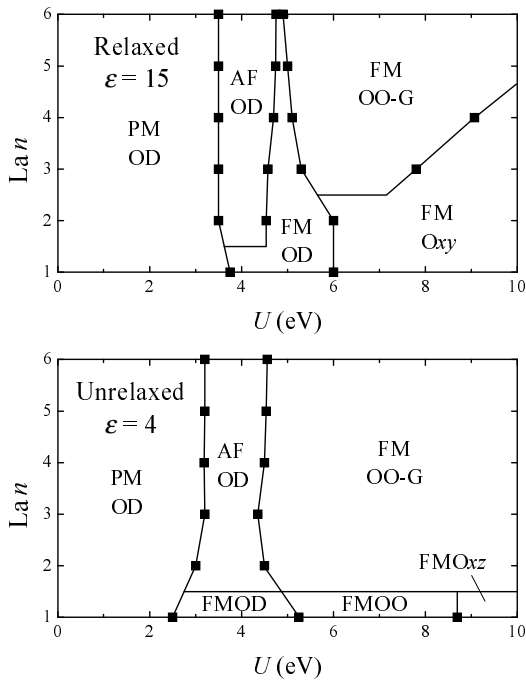}
\caption{Spin and orbital phase diagrams of realistic three-band model heterostructures 
as a function of the intraorbital Coulomb interaction $U$ with 
interorbital Coulomb interaction $U'=U-2J$ and exchange interaction $J=0.6$~eV 
and La layer thickness $n$ computed within the Hartree-Fock approximation. 
Upper panel: relaxed structure with $\varepsilon=15$ and appropriate band parameters obtained from LDA+$U$, 
lower panel: unrelaxed structure with $\varepsilon=4$. 
PM: paramagnetic, FM: ferromagnetic states, 
AF: antiferromagnetic state in which the magnetic moment alternates from plane to plane, 
OD: orbitally disordered state, 
OO-G: orbitally ordered state in which $xz$ and $yz$ orbitals alternate in $x,y$ and $z$ directions, 
O$xy(xz)$: orbitally ordered state in which $xy$ $(xz \, {\rm or} \, yz)$ occupancy is predominant, and  
OO: orbitally ordered state in which $xz$ and $yz$ orbitals alternate in $z$ direction. 
}
\label{fig:diagram}
\end{figure}

Finally, to investigate the changes due to the lattice relaxations, we compute the many-body phase 
diagrams for the relaxed and unrelaxed cases using the band parameters derived above and the methods 
of Ref.~\cite{Okamoto04a}. To avoid complexity coming from the interference between different La regions 
we use isolated heterostructures in which $n$ LaTiO$_3$ layers were sandwiched by a semi-infinite number 
of SrTiO$_3$ layers. Figure \ref{fig:diagram} compares our phase diagrams for relaxed and unrelaxed 
heterostructures. 

First we comment on features that are common to both relaxed and unrelaxed cases. 
The present calculations are an improvement over previously published \cite{Okamoto04a} 
Hartree-Fock calculations in that they allow for the possibility of a fully alternating 
(both within plane and from plane to plane) antiferro-orbital ordering state. We denote 
this phase as OO-G, and see that it is in fact the ground state over wide regions of the 
phase diagram in both cases. This phase is favored by strong correlations (large $U$) and 
electron densities near one \cite{Mizokawa95}. 
In thicker heterostructures, at the edge of the high density region, 
we expect this phase to be replaced by a ferro-orbital 
phase for reasons similar to those discussed in Ref.~\cite{Lin06}; this will be discussed in detail 
elsewhere \cite{Okamoto06}. 

The most striking {\it differences} between the phase diagrams result from the different density profiles. 
In particular the lower central-layer charge density and wider transition regions of the relaxed structure 
shift the phase boundaries to higher $U$ values, and disfavor the OO-G phase. 
The lattice relaxation also changes the local environment at each Ti site, 
but in the calculations presented here these effects cannot be isolated from the larger effects 
due to changes in screening. Insight into the importance of the symmetry breaking at 
fixed screening is gained by comparing 
Fig.~\ref{fig:diagram} to our previously published
Fig.~1 of Ref.~\cite{Okamoto04a}-1 and Fig.~2 of Ref.~\cite{Okamoto04a}-1. 
These figures show a phase diagram corresponding to similar charge densities, 
but with isotropic hoppings and no level splitting. 
Note that when the previous figure was constructed, only calculations with in-plane translation invariance 
were possible, the OO-G phase could not be studied and a higher energy layered orbital state ($xz$-$yz$)
was found at all thicknesses. 
We see that the effect of the symmetry breaking, larger hopping intensities for $d_{xy}$ band, 
is to replace this phase by the ferro-orbital O$xy$ for small $n$. 
We note that the O$xy$ phase is not observed in the unrelaxed case at $n=1$, even with the symmetry-broken
hopping parameters because, in the unrelaxed $n=1$ case, the central Ti layer (Ti$_{\pm0.5}$) density 
is too high. 
Instead, the system prefers other orbitally ordered states with narrower $xy$-plane band widths. 

To summarize, we have performed first-principles calculation on LaTiO$_3$/SrTiO$_3$ superlattices and 
investigated the effect of lattice relaxation on the charge screening and change in the model parameters. 
Our primary observation is a large polar distortion of TiO$_6$ octahedra at the near La region in which 
Ti and O ions are displaced, leading to screening which reduces the central layer charge density, substantially 
consistent with experiment \cite{Ohtomo02} and disfavoring staggered orbital orderings. In addition, for 
thin heterostructures at strong correlation, the symmetry breaking due to the distortion favors a novel 
uniform orbital ordering O$xy$ not found in bulk. 

We thank A. M. Rappe, C. Ederer and D. R. Hamann for helpful discussions. 
S.O. acknowledges the Materials Research Laboratory, UC Santa Barbara for hospitality. 
This research was supported by the DOE under Grant No. ER 46169 (A.M. and S.O.), by the NSF under
grant number CHE-0434567 (N.S.), and made use of MRL Central Facilities supported by the MRSEC Program 
of the National Science Foundation under award No. DMR05-20415.


\begin{thebibliography}{99}

\bibitem{Imada98} M.~Imada, A. Fujimori, and Y. Tokura, Rev. Mod. Phys. 
\textbf{70}, 1039 (1998).

\bibitem{Tokura00} Y. Tokura and N. Nagaosa, Science \textbf{288}, 462
(2000).

\bibitem{Izumi01} M. Izumi, Y. Ogimoto, Y. Konishi, T. Manako, M. Kawasaki,
and Y. Tokura, Mat. Sci. Eng. B \textbf{84}, 53 (2001) and references
therein.

\bibitem{Ohtomo02} A. Ohtomo, D. A. Muller, J. L. Grazul, and H. Y. Hwang,
Nature (London) \textbf{419}, 378 (2002).

\bibitem{Gariglio02}S. Gariglio, C. H. Ahn, D. Matthey, and J.-M. Triscone, 
Phys. Rev. Lett. {\bf 88}, 067002 (2002). 

\bibitem{Bozovic03}
I. Bozovic, G. Logvenov, M. A. J. Verhoeven, P. Caputo, E. Goldobin, T. H. Geballe, 
Nature (London) {\bf 422}, 873 (2003); 
I. Bozovic, G. Logvenov, M. A. J. Verhoeven, P. Caputo, E. Goldobin, and M. R. Beasley, 
Phys. Rev. Lett. {\bf 93}, 157002 (2004). 

\bibitem{Yamada04}H. Yamada, Y. Ogawa, Y. Ishii, H. Sato, M. Kawasaki, H. Akoh, and Y. Tokura,
Science {\bf 305}, 646 (2004). 

\bibitem{Varela04}M. Varela, S. D. Findlay, A. R. Lupini, H. M. Christen, A. Y. Borisevich, N. Dellby, 
O. L. Krivanek, P. D. Nellist, M. P. Oxley, L. J. Allen, and S. J. Pennycook, 
Phys. Rev. Lett. {\bf 92}, 095502 (2004). 

\bibitem{Schneider04}C. W. Schneider, S. Hembacher, G. Hammerl, R. Held, A. Schmehl, A. Weber, 
T. Kopp, and J. Mannhart, Phys. Rev. Lett. {\bf 92}, 257003 (2004). 

\bibitem{Pena05}V. Pe{\~ n}a, Z. Sefrioui, D. Arias, C. Leon, J. Santamaria, J. L. Martinez, 
S. G. E. te Velthuis, 
and A. Hoffmann, Phys. Rev. Lett. {\bf 94}, 057002 (2005). 

\bibitem{Stahn05}J. Stahn, J. Chakhalian, Ch. Niedermayer, J. Hoppler, T. Gutberlet, J. Voigt, 
F. Treubel, H-U. Habermeier, G. Cristiani, B. Keimer, and C. Bernhard, Phys. Rev. B {\bf 71}, 140509(R) (2005). 

\bibitem{Logvenov05}G. Yu. Logvenov, C. W. Schneider, J. Mannhart, and Yu. S. Barash, 
Appl. Phys. Lett. {\bf 86}, 202505 (2005).

\bibitem{Kavach05}J. J. Kavich, M. P. Warusawithana, J. W. Freeland, P. Ryan, X. Zhai, R. H. Kodama, 
J. N. Eckstein, cond-mat/0512158. 

\bibitem{Okamoto04a}S. Okamoto and A. J. Millis, Nature (London) \textbf{428}, 630 (2004); 
Phys. Rev. B {\bf 70}, 075101 (2004).

\bibitem{Okamoto04b}S. Okamoto and A. J. Millis, Phys. Rev. B {\bf 70}, 241104(R) (2004). 

\bibitem{Okamoto05}S. Okamoto and A. J. Millis, Phys. Rev. B {\bf 72}, 235108 (2005). 


\bibitem{Popovic05}Z. S. Popovic and S. Satpathy, Phys. Rev. Lett. {\bf 94}, 176805 (2005). 

\bibitem{Freericks04}J.~K.~Freericks, Phys. Rev. B {\bf 70}, 195342 (2004). 

\bibitem{Pavlenko05}N. Pavlenko and T. Kopp, Phys. Rev. B {\bf 72}, 174516 (2005). 







\bibitem{Bolechl94}P. E. Blochl, Phys. Rev. B {\bf 50}, 17953 (1994).

\bibitem{Kresse96}G. Kresse and J. Furthmuller, Phys. Rev. B {\bf 54}, 11169 (1996).
\bibitem{Kresse99}G. Kresse and D. Joubert, Phys. Rev. B {\bf 59}, 1758 (1999).

\bibitem{Liechtenstein95}A. I. Liechtenstein, V. I. Anisimov and J. Zaanen, 
Phys. Rev. B {\bf 52}, R5467 (1995). 

\bibitem{Mizokawa95}T. Mizokawa and A. Fujimori, Phys. Rev. B {\bf 51}, R12880 (1995). 

\bibitem{Czyzyk94}M. T. Czy{\.z}yk and G. A. Sawatzky, Phys. Rev. B {\bf 49}, 14211 (1994). 

\bibitem{Sai05}N. Sai, A. M. Kolpak, and A. M. Rappe, Phys. Rev. B {\bf 72}, 020101(R) (2005). 


\bibitem{Slater54}J.~C.~Slater and G.~F.~Koster, Phys. Rev. {\bf 94}, 1498 (1954). 

\bibitem{Lin06}C. Lin, S. Okamoto, A. J. Millis (in preparation). 

\bibitem{Okamoto06}S. Okamoto, A. J. Millis, and N. A. Spaldin (in preparation). 

\end{thebibliography}
\end{document}